# Ultrafast modulation of near-field heat transfer with tunable metamaterials

Longji Cui, Yong Huang,[a)] Ju Wang, and Ke-Yong Zhu

*School of Aeronautic Science and Engineering, Beijing University of Aeronautics and Astronautics, Beijing 100191, China*

## Abstract

We propose a mechanism of active near-field heat transfer modulation relying on externally tunable metamaterials. A large modulation effect is observed and can be explained by the coupling of surface modes, which is dramatically varied in the presence of controllable magnetoelectric coupling in metamaterials. We finally discuss how a practical picosecond-scale thermal modulator can be made. This modulator allows manipulating nanoscale heat flux in an ultrafast and noncontact (by optical means) manner.

Radiative heat transfer between surfaces in close proximity has attracted much attention since it has been predicted that the heat flux can break the Planck's blackbody radiation law.[1] This effect, due to the tunneling of evanescent waves, is pronounced only in the near-field, i.e., at separation of surfaces smaller than the characteristic thermal wavelength. Particularly, enhancement that is several orders of magnitude over blackbody limit occurs if the heat transfer involves thermally excited surface modes,[2-5] such as surface plasmon and phonon polaritons. Considerable theoretical effort has been devoted to understand the detailed mechanism of near-field heat transfer in a number of geometries[6-12] and materials.[13-19] Near-field thermal effect has been verified in recent experiments,[20-23] and holds promise for applications such as imaging,[24,25] energy conversion,[26-28] and noncontact localized heating.[29]

Despite these progress, relatively little attention has been focused on how to actively control this transfer. This has deep implications for applications in photonic thermal circuits and thermal management in nanoelectronics. Since near-field heat flux is intimately related to optical properties of materials, one possibility is to modulate heat flow by using materials whose properties can be tuned by external stimulus. To this end, several schemes have been put forward by employing phase-change materials,[30-32] graphene,[33,34] anisotropic structures,[35] as well as temperature dependent materials.[36,37] However, limited by properties of the applied materials, existing schemes involve only tuning the dielectric response. The potential to control heat flux with tunable magnetic and magnetoelectric coupling response has not been explored yet. In this paper, we present an alternative scheme for controlling near-field heat flux. It shows that, by using the size and sign of magnetoelectric coupling in metamaterials as controllable parameters, it is feasible to achieve a large heat flux modulation effect.

Metamaterials are artificially structured materials possessing many unusual properties, such as negative refraction,[38] magnetism up to optical frequencies,[39] and giant magnetoelectric coupling (chiral metamaterials).[40] These properties are gained from the unit-cells that replace atoms of natural materials as basic elements in light-matter interactions. Based on the flexibility of unit-cell design, tunable metamaterials are constructed by incorporating photo- or electric-active materials (e.g., semiconductors).[41-43] This makes it possible to switch the electromagnetic responses through photoexcitation or external voltage. Specifically, recent progress[44,45] has made metamaterials with tunable magnetoelectric coupling possible. More importantly, switching of different responses induced by photoexcitation can be achieved within very short time.[41,46] Therefore, this metamaterial has great potential to make an ultrafast thermal modulator.

Consider two chiral metamaterials (labeled 1 and 2), modeled as planar semi-infinite objects, separated by a vacuum gap of thickness $d$, and maintained in local thermodynamic equilibrium with temperature $T_1$ and $T_2$. The radiative heat flux across the gap can be calculated by fluctuational electrodynamics introduced by Rytov.[47] In this framework, both the far-field contribution and the near-field enhancement of heat transfer have been taken into account. In general, firstly the correlation function of stochastic current density is related to the system temperature by fluctuation dissipation theorem; then the fluctuational field is solved by Green's dyadic technique; finally the heat flux is obtained from the statistically averaged Poynting vector. For the case of planar parallel bodies considered here, the net near-field heat flux $q$ between the two surfaces is given by[19]

$$q = \int_0^\infty \frac{d\omega}{2\pi} [\Theta(\omega,T_2) - \Theta(\omega,T_1)]$$
$$\times \left\{ \int_{k_\parallel < k_0} \frac{dk_\parallel}{2\pi} k_\parallel \text{Tr}\left[(\mathbf{I} - \mathbf{R}_2^* \cdot \mathbf{R}_2) \cdot \mathbf{D}_{12} \cdot (\mathbf{I} - \mathbf{R}_1 \cdot \mathbf{R}_1^*) \cdot \mathbf{D}_{12}^*\right] \right. \quad (1)$$
$$\left. + \int_{k_\parallel > k_0} \frac{dk_\parallel}{2\pi} k_\parallel \text{Tr}\left[(\mathbf{R}_2^* - \mathbf{R}_2) \cdot \mathbf{D}_{12} \cdot (\mathbf{R}_1 - \mathbf{R}_1^*) \cdot \mathbf{D}_{12}^*\right] e^{-2|\gamma|d} \right\}$$

where $\Theta(\omega,T) = \hbar\omega/(\exp(\hbar\omega/k_B T) - 1)$ is the average energy of Planck's oscillator, $k_0 = \omega/c$, $k_\parallel$ and $\gamma = \sqrt{k_0^2 - k_\parallel^2}$ are, respectively, parallel and normal component of vacuum wavevector. Tr is the trace operator, the symbol * denotes hermitian conjugation, and $\mathbf{D}_{12} = (\mathbf{I} - \mathbf{R}_1 \cdot \mathbf{R}_2 e^{-2|\gamma|d})^{-1}$ with $\mathbf{R}_i$ the reflection matrix defined as

$$\mathbf{R}_i = \begin{bmatrix} r_i^{ss}(\omega,k_\parallel) & r_i^{sp}(\omega,k_\parallel) \\ r_i^{ps}(\omega,k_\parallel) & r_i^{pp}(\omega,k_\parallel) \end{bmatrix} \quad (i=1,2), \quad (2)$$

where $r_i^{ab}$ gives the ratio of the reflected waves with *b*-polarization to the incident waves with *a*-polarization, and *s* (*p*) corresponds to transverse magnetic (electric) polarization. For isotropic media, the off-diagonal elements in Eq. (2) vanish and the diagonal elements reduce to the well-known Fresnel coefficient in terms of electric permittivity $\varepsilon$ and magnetic permeability $\mu$. However for chiral media, the constitutive relation has the form[48] $D = \varepsilon_0 \varepsilon E + i\kappa H/c$ and $B = \mu_0 \mu H - i\kappa E/c$, where $c = 1/\sqrt{\varepsilon_0 \mu_0}$ is the vacuum light speed, and $\kappa$ is chirality parameter characterizing the magnetoelectric coupling response. In this case, both the diagonal and off-diagonal elements in the reflection matrices become non-zero due to magnetoelectric coupling effect. Assuming the incidence of the electromagnetic waves is from vacuum to chiral media, the reflection coefficients can be explicitly written as[48]

$$r_i^{ss} = -\frac{(\eta_0^2 - \eta^2)(\xi_+ + \xi_-) + 2\eta_0\eta(\xi_+\xi_- - 1)}{(\eta_0^2 + \eta^2)(\xi_+ + \xi_-) + 2\eta_0\eta(\xi_+\xi_- + 1)}, \tag{3a}$$

$$r_i^{pp} = \frac{(\eta_0^2 - \eta^2)(\xi_+ + \xi_-) - 2\eta_0\eta(\xi_+\xi_- - 1)}{(\eta_0^2 + \eta^2)(\xi_+ + \xi_-) + 2\eta_0\eta(\xi_+\xi_- + 1)}, \tag{3b}$$

$$r_i^{sp} = -r_i^{ps} = \frac{2i\eta_0\eta(\xi_+ - \xi_-)}{(\eta_0^2 + \eta^2)(\xi_+ + \xi_-) + 2\eta_0\eta(\xi_+\xi_- + 1)}, \tag{3c}$$

where $\eta_0 = \sqrt{\mu_0/\varepsilon_0}$, $\eta_i = \sqrt{\mu_0\mu_i/\varepsilon_0\varepsilon_i}$, and $\xi_\pm = \gamma_\pm/n_\pm\gamma$ in which $n_\pm = \sqrt{\varepsilon_i\mu_i} \pm \kappa_i$ and $\gamma_\pm = \sqrt{n_\pm^2 k_0^2 - k_\parallel^2}$ are, respectively, the refraction indices and normal wavevector components of the two circular polarized waves in chiral media.

Now we proceed to demonstrate how near-field heat flux can be modulated by using the dynamic characteristic of magnetoelectric coupling in metamaterials. The constitutive parameters of chiral media with single resonance are written in a general form[49]

$$\varepsilon(\omega) = 1 - \frac{\Omega_\varepsilon \omega_\varepsilon^2}{\omega^2 - \omega_\varepsilon^2 + i\Upsilon_\varepsilon\omega}, \tag{4a}$$

$$\mu(\omega) = 1 - \frac{\Omega_\mu \omega^2}{\omega^2 - \omega_\mu^2 + i\Upsilon_\mu\omega}, \tag{4b}$$

$$\kappa(\omega) = \frac{\pm\Omega_\kappa \omega\omega_\kappa}{\omega^2 - \omega_\kappa^2 + i\Upsilon_\kappa\omega}, \tag{4c}$$

Where $\omega_\varepsilon$, $\omega_\mu$, and $\omega_\kappa$ denote the resonance frequencies of the dielectric, magnetic, and chirality responses; $\Upsilon_\varepsilon$, $\Upsilon_\mu$, and $\Upsilon_\kappa$ are the damping factors; $\Omega_\varepsilon$, $\Omega_\mu$, and $\Omega_\kappa$ account for the resonance strength of the permittivity, permeability, and chirality, respectively. For chiral metamaterials, the dielectric, magnetic, and chirality resonances generally have the same frequency, $\omega_0 = \omega_\varepsilon = \omega_\mu = \omega_\kappa$. Moreover, the second law of thermodynamics requires that the lossy part of the constitutive parameters must satisfy the inequality,[50] $\text{Im}^2(\kappa) \leq \text{Im}(\varepsilon)\text{Im}(\mu)$, therefore, there exists an upper limit for $\Omega_\kappa$. For the convenience of numerical analysis, we assume the same damping factor, $\Upsilon = \Upsilon_\varepsilon = \Upsilon_\mu = \Upsilon_\kappa$, the limit value can thus be obtained by substituting Eqs. (4a)-(4c) into the inequality, i.e., $\Omega_\kappa = \sqrt{\Omega_\varepsilon \Omega_\mu}$.

In what follows, two metamaterial designs with distinct tunability are considered: the design in Ref. 44 allowing switching chiral metamaterial between its initial handedness and the opposite handedness, and the design in Ref. 45 enabling continuous modulation of chirality. It should be noted that these metamaterials are designed specifically for electromagnetic responses in THz frequency region. Since the response frequency of a metamaterial scales inversely with its unit-cell geometrical dimension up to optical frequency (i.e., the size-scaling rule),[51-53] it is expected that these designs can be miniaturized to possess responses in the infrared that is of interest for near-field radiative heat transfer. In principle, the size-scaling rule keeps valid as long as the metal in the unit-cell behaves as ideal metal. According to Ref. 53, the minimum dimension of the size-scaling regime for effective miniaturization of the unit-cell is approximately 55 nm. Here, it must be pointed out that since this minimum value is for a specific design of metamaterial (i.e., split ring resonator), a rigorous numerical simulation should be carried out to determine the separation at which the size-scaling rule breaks down for other metamaterial designs.

We first consider a setup including two metamaterials with the constitutive parameters, $\varepsilon_1 = \varepsilon_2$, $\mu_1 = \mu_2$, $\kappa_1 = \kappa_2$ (without control) and $\kappa_1 = -\kappa_2$ (under control). To quantify the heat flux modulation induced by the chirality reversion, we define the relative difference of heat transfer coefficients between the two configurations as $\mathrm{M} = (h_O - h_C)/h_O$, where $h = \lim_{T_1 \to T_2 = T} [q/(T_2 - T_1)]$, and the subscript $O$ and $C$ corresponds to the original and controlled state, respectively. Without loss of generality, the parameters are taken as $\omega_0 = 3 \times 10^{14}\,\mathrm{rad/s}$, $\Upsilon = 0.01\omega_0$, $\Omega_\varepsilon = 2$, $\Omega_\mu = 0.5$, and $T_2 - T_1 \ll T_2 = 300\mathrm{K}$. Figure 1(a) displays modulation M as a function of dimensionless separation distance $\bar{d} \equiv d/\lambda_0$ ($\lambda_0 \equiv (2\pi c)/\omega_0$ is the resonance wavelength) for different sizes of chirality. It shows that modulation of heat transfer exists since the heat transfer coefficient in the case of two identical metamaterials ($\kappa_1 = \kappa_2$) is always larger than that in the case of opposite chirality ($\kappa_1 = -\kappa_2$). Moreover, the modulation favors configurations with strong chirality and the maximum modulation shown in Fig. 1(a) is found to be 0.7.

In order to understand the physical origin of heat flux modulation, we examine the distribution of near-field heat flux in the $(k_\parallel, \omega)$ plane, which is dictated by the integrand of Eq. (1). As shown in Fig. 1(b) and 1(c), for both configurations, heat flux is mainly transferred through channels in the vicinity of the surface mode resonance frequencies ($\omega = 1.08\omega_0$ and $\omega = 1.60\omega_0$ for the system considered here). This indicates that when the two parallel surfaces are closely spaced in the near field, heat transfer is dominated by the coupling of surface modes at the resonance frequencies. Since the dispersion relation of the surface modes corresponds to the pole of the reflection coefficients in Eq. (3), the resonances can be found by evaluating the asymptotic behavior of these coefficients. Indeed, in the limit $k_\parallel \gg k_0$, $r_{ss} \approx [(\varepsilon+1)(\mu-1) - \kappa^2]/\sigma$, $r_{pp} \approx [(\varepsilon-1)(\mu+1) - \kappa^2]/\sigma$, and $r_{sp} \approx -2\kappa i/\sigma$ with $\sigma = (\varepsilon+1)(\mu+1) - \kappa^2$. The surface mode resonances are therefore exhibited at frequencies where $(\varepsilon+1)(\mu+1) = \kappa^2$. It should be noted here that the resonance frequencies are independent of the sign of $\kappa$. However, the coupling efficiency of the surface modes shows significant difference as sign of $\kappa$ is reversed. For the configuration of opposite $\kappa$, we observe that the coupling of surface modes fades out at a cutoff value $k_\parallel^c \approx 40 k_0$ for $\omega = 1.08\omega_0$, and $k_\parallel^c \approx 50 k_0$ for $\omega = 1.60\omega_0$. In contrast, more coupling modes ($k_\parallel^c \approx 50 k_0$ and $65 k_0$ for two resonances, respectively) and stronger coupling strength is found to contribute to heat transfer for the configuration of same handedness.

It is clear from the above analysis that two discrete levels of near-field heat flux can be achieved by only switching the handedness of metamaterials, now we turn to investigate the heat flux modulation using dynamical metamaterials that allow tuning $\kappa$ continuously. In Fig. 2(a), we plot the logarithmic contour of heat flux between two metamaterials with the constitutive parameters, $\varepsilon_1 = \varepsilon_2$, $\mu_1 = \mu_2$, and $\Omega_{\kappa 1}, \Omega_{\kappa 2} \in [-1, 1]$. It shows that at fixed separation distance considered here ($\bar{d} \approx 0.01$), different combinations of $\Omega_{\kappa 1}$ and $\Omega_{\kappa 2}$ lead to a wide range modulation of heat flux. The maximal heat flux occurs around $\Omega_{\kappa 1} = \Omega_{\kappa 2} = 0.1$ (P2 in Fig 2(a)) and is about 7% larger than the hear flux in the absence of magnetoelectric coupling in materials. It is also found that the maximal hear flux is more than one order larger than the minimal heat flux which occurs in the thermodynamic limit of the chiral media [P4, $\Omega_{\kappa 1} = -\Omega_{\kappa 2} = 1.0$]. The pattern of heat flux modulation shown on the contour and the large heat transfer contrast can be attributed to the contribution of coupled surface modes in different configurations. We note that, symmetrical or anti-symmetrical configurations ($\kappa_1 = \kappa_2$ or $\kappa_1 = -\kappa_2$) tend to generate a large heat flux between the surfaces. The underlying mechanism is that for such configurations, surface mode resonances are excited on both material surfaces at identical frequencies, and thus lead to more efficient coupling of surface modes at the resonances. On the contrary, for non-symmetrical configurations, surface modes on the two surfaces move out of resonance, resulting in weak coupling. As confirmed in Fig. 2(b), where the heat flux spectral variation for different points (from P1 to P4) is plotted, the heat transfer contrast in the figure is mainly due to the rapidly falling off of $h$ at the surface mode resonance frequencies.

In the following we discuss in more detail the possibility to make a practical heat flux modulator. Our theoretical predictions are based on effective medium approximation (EMA) that the metamaterial is regarded as homogeneous medium characterized by effective constitutive parameters shown in Eqs. (4a)-(4c). In fact, the EMA can be used as long as the wavelength of the important electromagnetic fluctuation, which is on the order of the separation distance between surfaces, is larger than the atomic dimension of the medium,[54] i.e., $d > l$, where $l$ is the unit-cell dimension. Moreover, one can also consider that the a nonlocal model is necessary if the parallel wave vectors of evanescent modes are on the order of $\pi/l$. Since the exponential in the transmission coefficient in Eq. (1) sets a cut-off wave vector of evanescent modes contributing to near-field heat transfer which is $\sim 1/d$, one finds that a local effective medium description is permissible for $d > l/\pi$. Hence, it can be seen that for the given unit-cell dimension considered here, the separation distances set at $d > 60 nm$ can be viewed as appropriate. However, the EMA should be cautiously applied in small separation distances and a precise minimum separation at which the EMA is invalid can only be obtained through numerical simulation for a concrete material.

On the other hand, note that thermal modulation is pronounced at $d \lesssim 0.01\lambda_0$, indicating that the metamaterial should possess a large wavelength-to-structure ratio ($\lambda_0/a > 100$) to ensure a well defined EMA. In reality, some specific metamaterials with very compact inner structure (e.g., the Swiss roll design[40] with $\lambda_0/a$ larger than a few hundred) can be good candidates for this purpose. However, it should also be noted that due to its compact structure, the suggested Swiss roll design cannot guarantee neither the electromagnetic response in the infrared nor a separation distance smaller than a hundred nanometers at which the EMA is valid. In reality, for practical near-field thermal application, the present scheme is still limited by current metamaterial design methods. In addition, the fabrication of complex metamaterials structures on micron/nanometer scale remains to be a challenging problem.

To illustrate the potential of tunable metamaterial for fast heat flux modulation, we compare our scheme with the reported schemes using phase-change material[30,31] and graphene-based structure.[33] For photoexcitation controlled metamaterials, the tunability of the electromagnetic responses is accomplished by generating a high concentration of free charge carrier (and thus change the electric conductivity of the metamolecules) using an optical pulse. The pump fluence of the pulse enables to modulate the strength of the response. The switching time between different responses is determined by the carrier lifetime of the photo-active materials integrated into the metamolecules. It has been shown that by appropriately engineering the materials with short carrier lifetime, the switching time can be minimized to be as short as a few picoseconds.[46]

Thus, it is seen that an ultrafast thermal modulator can be made. Moreover, the heat flux control can be either continuous or discrete, depending on specific design of the metamaterials (as shown in our calculation). In comparison, heat flux modulation using the switching of two discrete states (crystal-amorphous or metal-insulator) of phase-change materials is on the order of a few nanoseconds, which is several orders of magnitude slower than the present scheme; instead the high carrier mobility of graphene that can be tuned by gating enables a high operating frequency up to a hundred gigahertz, comparable to the switching frequency of dynamic metamaterials. Furthermore, after a limited number of cycles, thermal modulator made with phase-change materials ceases to be reversible since the heating pulse causes incomplete state transition, and eventually leading to indistinguishable dielectric property between different phases.[55] On the contrary, reversibility of the present scheme relies on the carrier recombination that can be completely recovered to the initial state transiently following the optical pulse. Therefore, this suggests a different mechanism for improving cycle endurance of thermal modulator.

In conclusion we show that modulation of near-field heat transfer can be realized by means of metamaterials which can be externally tuned for obtaining desired electromagnetic responses. We show that, by using the size and sign of magnetoelectric coupling response as controllable parameters, a large heat flux modulation effect is feasible. It must be pointed out that our purpose of this Letter is to reveal the possibility of active nanoscale thermal modulation that can be operated ultrafast and by optical means, rather than give a complete account. Therefore, there exist plenty of other possibilities for obtaining noncontact and ultrafast thermal modulation based on the concept proposed here, such as using optically controlled semiconductors and even conceiving of the schemes working effectively in the ultra-near-field regime where the EMA breaks down. Along with the present scheme, we believe novel metamaterials that are specifically designed for near-field thermal application are necessary. On the other hand, for simplicity we consider the simple parallel planes and ideal metamaterial responses, further practical realization of this scheme can be extend to more general configurations and complex metamaterial properties, which can benefit from the great flexibility in engineering geometrical and optical properties of metamaterials.

## Acknowledgements

Y. H. gratefully acknowledges support from the Fundamental Research Funds for the Central Universities (No.YWF-11-03-Q-028).

a)Electronic mail: huangy@buaa.edu.cn

**Figure Caption**

FIG. 1. (Color online) (a) Near-field heat flux modulation effect due to chirality reversion, defined as relative difference of heat transfer coefficients between two chiral metamaterials with same handedness ($\kappa_1 = \kappa_2$) and opposite handedness ($\kappa_1 = -\kappa_2$), as a function of separation distance. The contour plot (a.u.) of the integrand of Eq. (1) in the $(k_\parallel, \omega)$ plane is displayed for (b) the configuration of same handedness, and (c) the configuration of opposite handedness, at $\Omega_\kappa = 0.9$ and $d/\lambda_0 = 0.01$.

FIG. 2. (Color online) (a) Logarithmic contour plot of heat flux between two tunable metamaterials with $\Omega_\kappa$ individually varying in the interval $[-1,1]$ at $d/\lambda_0 = 0.01$. Different combinations of $\Omega_{\kappa 1}$ and $\Omega_{\kappa 2}$ are marked from P1 to P4 (P1 is the case with no magnetoelectric coupling; P2 and P4 denotes the maximal and minimal heat flux respectively). (b) Spectral variation of near-field heat transfer coefficient for P1 to P4 marked in (a).

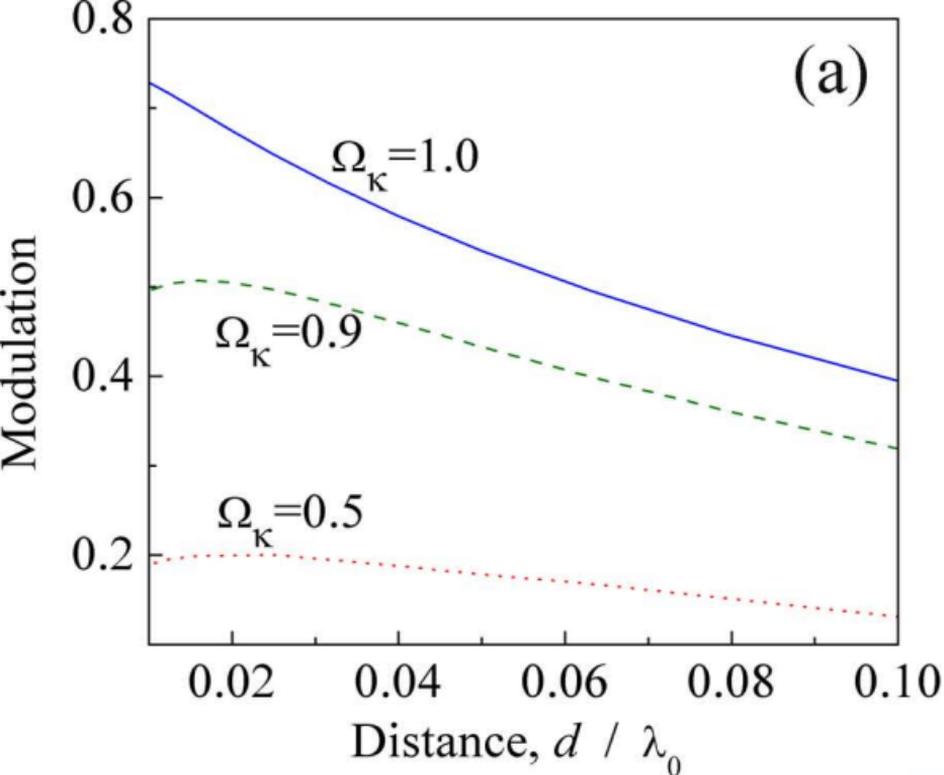

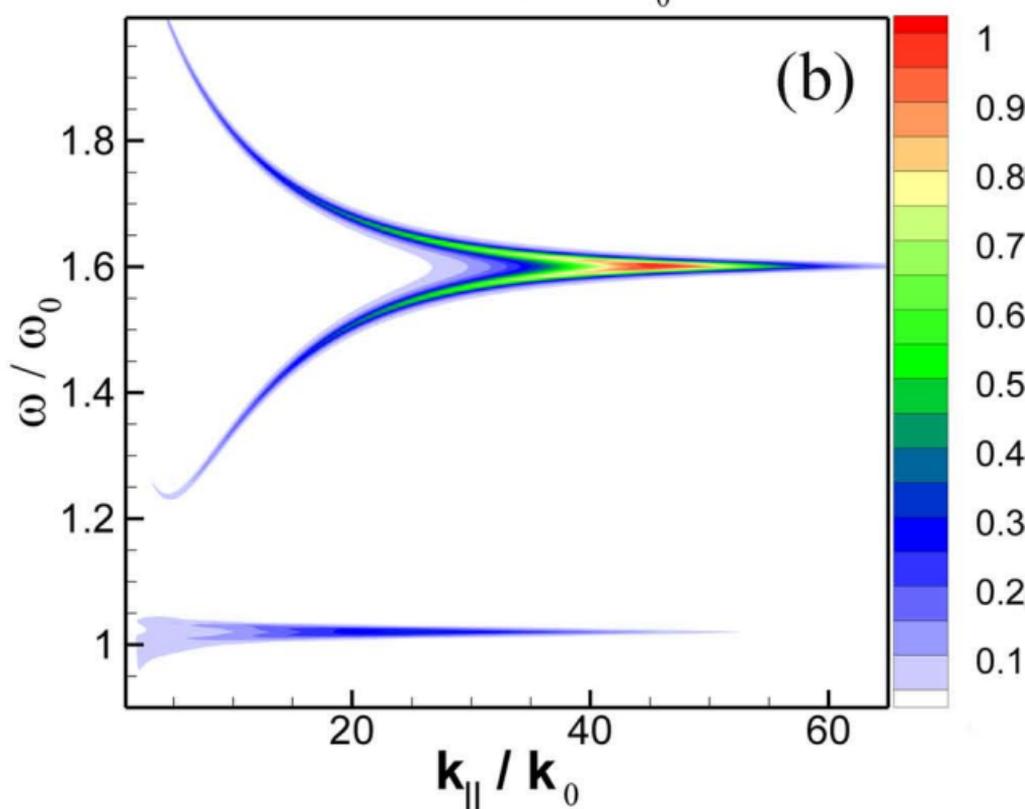

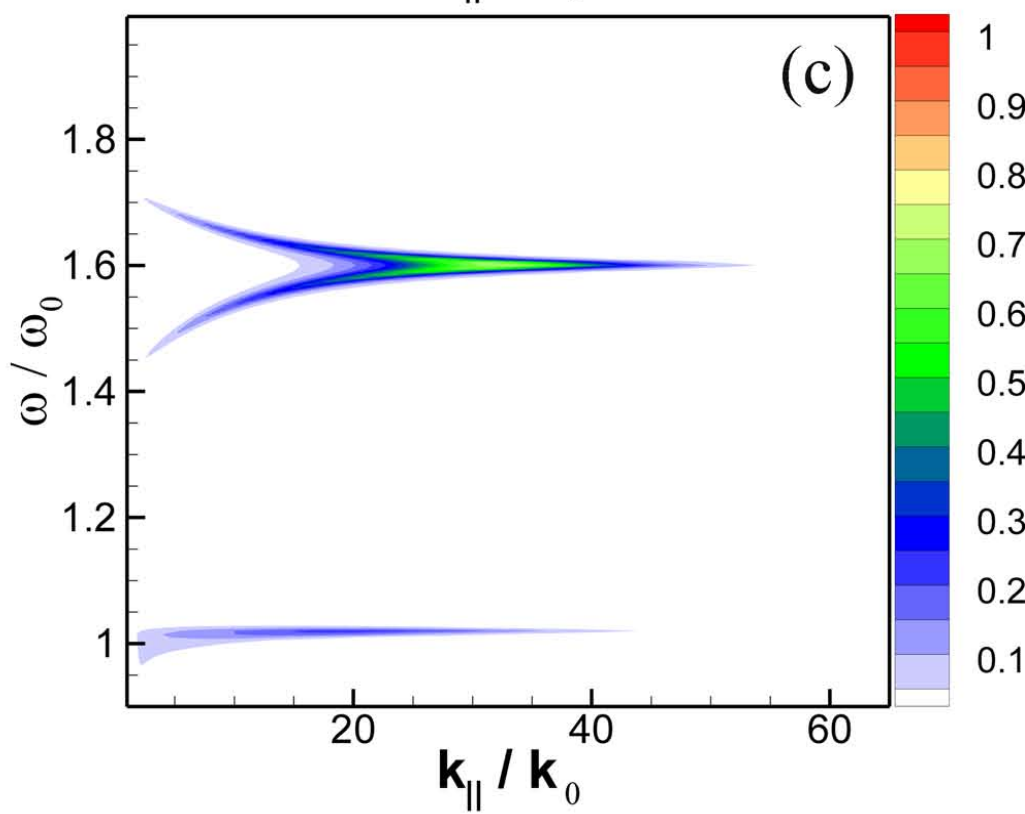

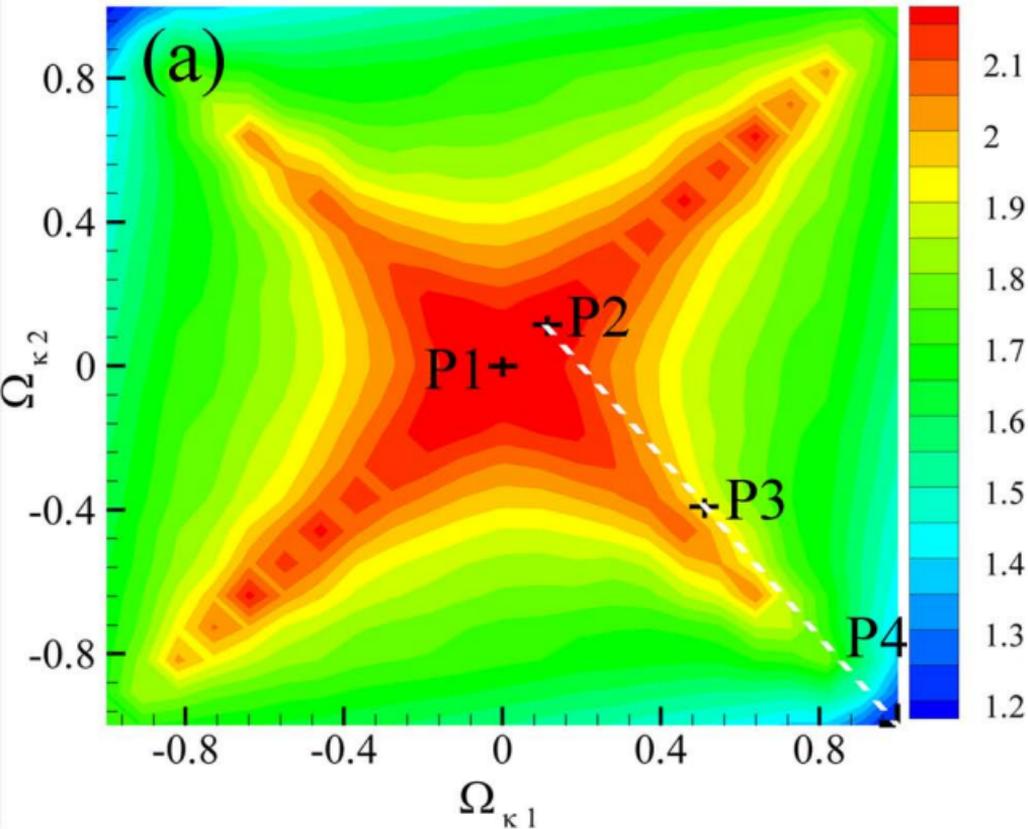